*Nuclei in the Cosmos 2000, Aarhus, Denmark, June 27 - July 1, 2000*

News from the p-process: is the s-process a troublemaker?

M. Rayet[a] [*], V. Costa[b] [†] and M. Arnould[a]

[a]Institut d'Astronomie et d'Astrophysique, Université Libre de Bruxelles, Belgium

[b]Dipartimento di Fisica e Astronomia dell'Università degli studi di Catania, and INFN-LNS, Catania, Italy

The most detailed calculations of the p-process call for its development in the O/Ne layers of Type II supernovae. In spite of their overall success in reproducing the solar system content of p-nuclides, they suggest a significant underproduction of the light Mo and Ru isotopes. On grounds of a model for the explosion of a 25 $M_\odot$ star with solar metallicity, we demonstrate that this failure might just be related to the uncertainties left in the rate of the $^{22}$Ne$(\alpha, n)^{25}$Mg neutron producing reaction. The latter indeed has a direct impact on the distribution of the s-process seeds for the p-process.

## 1. THE P-PROCESS IN TYPE II SUPERNOVAE

The most successful p-process models available to-date call for the synthesis of the stable neutron-deficient nuclides heavier than Fe (p-nuclei) in the O/Ne layers of Type II supernovae (SNII) ([13], hereafter RAHPN). Extensive calculations have been performed in RAHPN on the basis of stellar models which follow the evolution of helium stars from the beginning of helium burning up to, and including, the supernova explosion [10]. SNII explosions have been considered for stars with (approximately) 13, 15, 20 and 25 $M_\odot$ main sequence masses. The p-nuclei are produced in layers with explosion temperatures peaking in the $(1.8-3.3)\times 10^9$ K range, referred to as the P-Process Layers (PPLs). For each stellar mass we characterize the abundance of a p-nucleus $i$ by its mean overproduction factor $\langle F_i \rangle = \langle X_i \rangle / X_{i,\odot}$, $X_{i,\odot}$ being its solar mass fraction [1], and

$$\langle X_i \rangle = \frac{1}{M_p} \sum_{n \geq 1}(X_{i,n} + X_{i,n-1})(M_n - M_{n-1})/2, \qquad (1)$$

where $X_{i,n}$ is the mass fraction of nuclide $i$ at the mass coordinate $M_n$, $M_p = \sum_{n \geq 1}(M_n - M_{n-1})$ is the total mass of the PPLs, the sum running over all the PPLs ($M_0$ corresponds to the bottom layer). An overproduction factor averaged over all 35 p-nuclei is calculated as $F_0 = \sum_i \langle F_i \rangle /35$, and is a measure of the global p-nuclide enrichment in the PPLs. So, if the computed p-nuclei abundance distribution were exactly solar, the normalized mean overproduction factors $\langle F_i \rangle /F_0$ would be equal to unity for all $i$. Figure 1a shows the range of variation of the latter quantities over the considered stellar mass range. Integrated yields are calculated by a convolution of the yields $M_p \langle X_i \rangle$ for a star of mass $M$ with the Initial Mass Function (IMF) $\xi(M) \propto M^{-2.7}$. The corresponding, IMF averaged, factors $\langle F_i \rangle / F_0$ are shown as full squares.

[*]M.R. is Research Associate of the FNRS (Belgium).

[†]V.C. thanks the E.U. for financial support of the Ph.D. in Physics at the University of Catania through the FSE.

Figure 1a shows that about 60% of the IMF averaged p-nucleus overproductions in SNII fit the solar system composition within a factor 3 and that the calculated overproductions do not depend drastically on the stellar mass. However, the SNII scenario suffers from some shortcomings. From the discussions of RAHPN (and of earlier works referred to in that paper) it transpires that the most embarassing one is the severe underproduction of the light Mo ($^{92-94}$Mo) and Ru ($^{96-98}$Ru) isotopes.

The sensitivity of the p-process overabundances to some changes in the input physics has been checked in RAHPN for the explosion of a 25 $M_\odot$ star. For example, the PPLs, and consequently the predicted p-process abundances, are sensitive to the key $^{12}C(\alpha,\gamma)^{16}O$ reaction rate used to calculate the pre-supernova models. The 25 $M_\odot$ star $\langle F_i \rangle / F_0$ factors of Fig. 1a were obtained with the 1985 value of this rate [5] (model 85). They are compared in Fig. 1b to those obtained with the use of the 1988 [6] rate (model 88). The production of p-nuclei has also been examined in model 88 but with a deposited explosion energy of $1.5\,10^{51}$ erg instead of the standard value $1.0\,10^{51}$ erg [10] (model 88HE). The same qualitative agreement with the p-nuclei solar system composition is obtained in the 3 models, along with the persistent underproduction of the Mo and Ru p-isotopes.

Finally, the effect on $\langle F_i \rangle / F_0$ of uncertainties in the reaction rates used in the p-process reaction network is shown in Fig. 1c. The overabundances obtained in RAHPN with model 85 are compared with the ones predicted in the same PPLs but with updated reaction rates. The latter calculation uses the rates predicted by the NACRE compilation [2] for charged particle captures by nuclei up to $^{28}$Si and the rates predicted by the new Hauser-Feshbach code MOST [8] for heavier targets [3], except for the experimentally-based neutron capture rates provided by Beer et al. [4]. While the production of some of the heavier p-nuclei appears to be quite sensitive to changes in the Hauser-Feshbach rates, the observed underproduction of Mo and Ru is again very robust to changes in the input physics.

Exotic solutions have been proposed to remedy this underproduction problem, calling in particular for accreting neutron stars or black holes (e.g. [15]). The level of the contribution of such sites to the solar system content of the p-nuclides is impossible to assess in any reliable way.

## 2. THE SEED ABUNDANCES

Since in the SNII scenario the p-nuclei are mainly produced by the photodisintegrations of pre-existing s-nuclide seeds produced in the He-burning core of massive progenitor stars, the initial distribution of s-nuclei is a key ingredient of the p-process calculations.

For the p-process calculations considered in Fig. 1a, the seed abundances are taken -consistently- from the core He burning s-process calculated for each stellar mass $M$ by Prantzos et al. [12], so that differences in the calculated p-process abundances reveal

---

[3]The NACRE and MOST rates are available in the Brussels Nuclear Astrophysics Library (http://www-astro.ulb.ac.be)

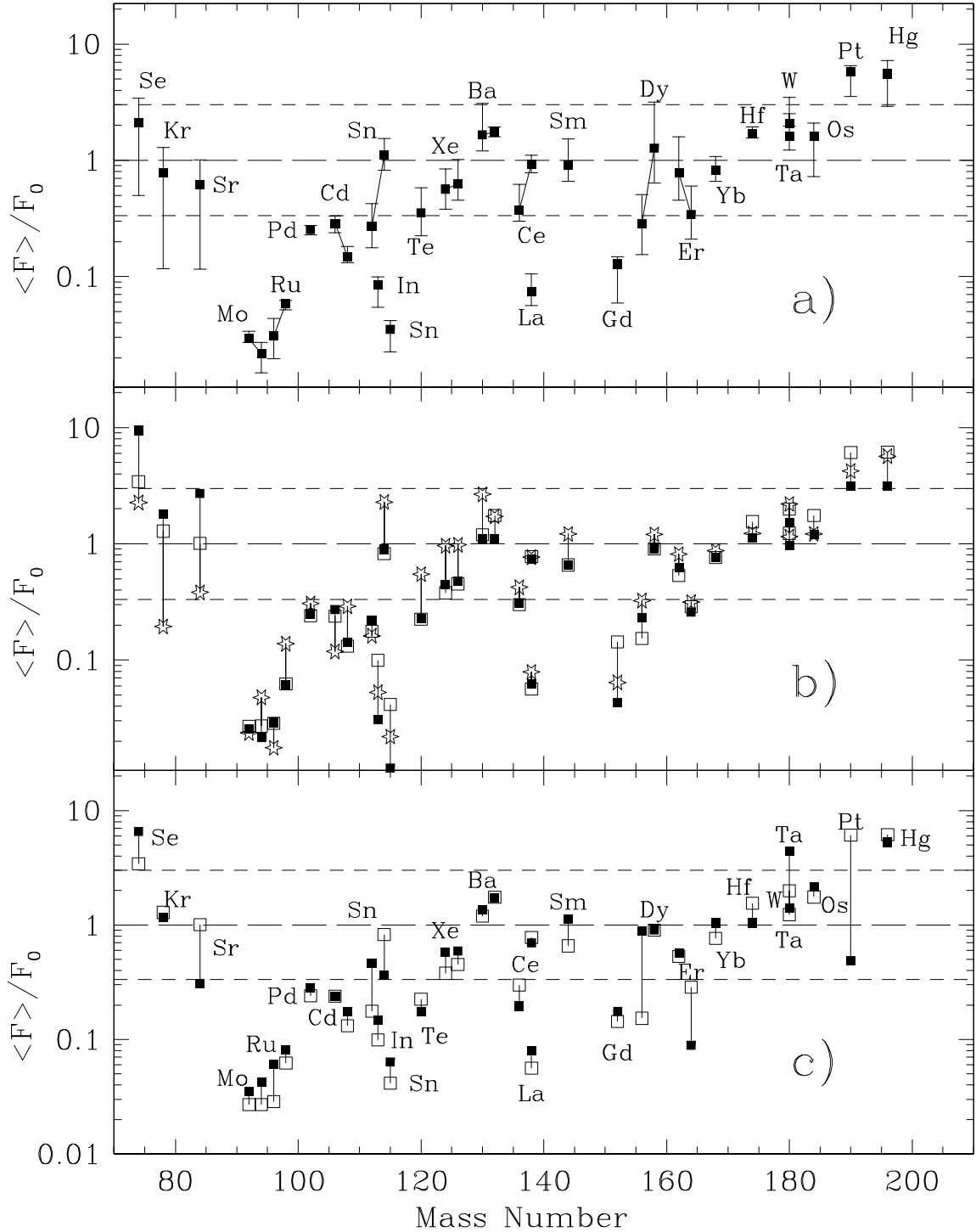

Figure 1. a) range of variation of $\langle F_i \rangle / F_0$ for the $13 - 25$ M$_\odot$ model stars and the IMF averaged values (full squares); b) values of $\langle F_i \rangle / F_0$ calculated for models 85 (open

both differences in the PPLs *and* in the seed abundances. On the other hand, in the comparisons made in Fig. 1b and 1c, the same seed abundances are -inconsistently- used throughout (equal to those obtained in [12] for a 25 $M_\odot$ star with the $^{12}$C$(\alpha,\gamma)^{16}$O rate from [6]) in order to isolate the effect due to the considered changes in astrophysical and nuclear physics inputs.

Although a systematic study of the influence of the seed nuclei distribution on the shaping of the calculated p-process abundances is still lacking, it has been emphasized many times over the last decade that the problem of the underproduction of the Mo and Ru isotopes might just be due to some misrepresentation of the production in the He-burning core of massive stars of the s-nuclide seeds for the p-process (e.g. RAHPN, [3]). In [7] we scrutinize the latter, 'non-exotic', solution in a quantitative way by duly taking into account the uncertainties that still affect the rate of the $^{22}$Ne$(\alpha,n)^{25}$Mg reaction, as they appear in the NACRE compilation of reaction rates [2]. Clearly, these uncertainties in the key neutron producer in conditions obtained during central He burning in massive stars have a direct impact on the predicted abundances of the s-nuclide seeds for the p-process, as already analyzed quantitatively by Meynet & Arnould [11].

## 3. THE EFFECT OF THE $^{22}$Ne($\alpha$,n)$^{25}$Mg RATE ON THE SEED DISTRIBUTION

For temperatures of about $2 - 3 \times 10^8$ K at which the s-process typically develops during core He burning in massive stars (e.g. [14]), the NACRE upper limit on the $^{22}$Ne$(\alpha,n)^{25}$Mg rate is 50-500 times larger than the 'adopted' value (see [2] for details). In order to quantify the consequences of this situation for the predicted abundance distribution of the s-nuclide seeds for the p-process, and ultimately for the p-nuclide yields themselves, we perform nucleosynthesis calculations for five different rates ranging from the NACRE adopted value to its upper limit. These rates, labelled $R_i$ ($i$=1 to 5) in the following, range from the NACRE "adopted" values ($R_1$) to the NACRE "upper" values ($R_5$) in a geometric progression ($R_3$ is thus the geometrical mean between $R_1$ and $R_5$). They are used in the 25 $M_\odot$ model 85 referred to above to calculate the abundances of the s-process nuclides at the end of core He burning. The results are shown in Fig. 2 for the s-only nuclides. Use of $R_1$ leads to the classical 'weak' s-process component pattern (e.g. [14]), exhibiting a decrease of the overproduction (with respect to solar) of the s-nuclides by a factor ranging from ≈100 to about unity when the mass number $A$ increases from about 70 to 100. In the heavier mass range, the s-process 'main component' supposed to originate from low- or intermediate-mass stars takes over. This 'canonical' picture changes gradually with an increase of the $^{22}$Ne$(\alpha,n)^{25}$Mg rate, more $^{22}$Ne having time to burn, releasing more neutrons, before He exhaustion in the core. The direct result of this is a steady increase of the overproduction of heavier and heavier s-nuclides. For example, with the extreme $R_5$ rate, the overproduction factor increases from $10^3$ to $10^4$ for $A$ varying from about 70 to 90, before decreasing to a value around unity for $A \approx 150$ only.

At first sight, it might be felt that the s-process abundance distributions obtained with large enough $R_i$ values exhibit some unwanted or embarrassing features. One of these concerns the underproduction of the $A \approx 70 - 76$ s-nuclides relative to the $A \approx$

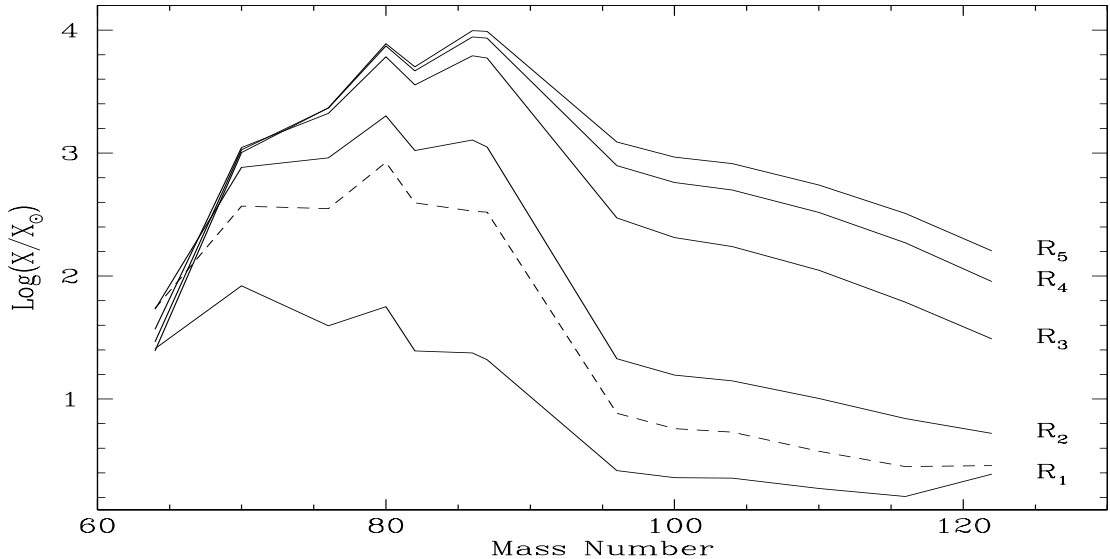

Figure 2. Distribution of the abundances, normalized to solar values, of the s-only nuclides at the end of core He burning in the considered 25 $M_\odot$ model star, for the $^{22}$Ne$(\alpha,n)^{25}$Mg rates $R_i$ (i = 1 to 5) defined in the text, all the other ingredients of the model being kept unchanged [10]. The dashed line is the distribution adopted by RAHPN for their p-process calculations

80 − 90 ones. Another one relates to the fact that a more or less substantial production of heavy s-nuclides (like in the Ba region) would screw up the pattern of the s-process main component ascribed to lower-mass stars. In our opinion, none of these predictions can really act as a deterrent to $^{22}$Ne$(\alpha,n)^{25}$Mg rates substantially in excess of $R_1$. On the one hand, the absence of ab initio self-consistent calculations of the s-process in low- and intermediate mass stars does not allow at this time to predict the exact shape of the main s-process component which is classically assigned to these stars. As a consequence, a contribution to the main component by massive stars cannot be excluded, even if it may disturb some traditional views on the subject. On the other hand, the reduction of the light s-process nuclide production by massive stars could well be compensated by their increased synthesis by some low- or intermediate-mass stars when rates larger than $R_1$ are considered [9]. The classical $^{80}$Kr overproduction problem found in the massive star s-process (e.g. [14]) could also be eased with increased $^{22}$Ne$(\alpha,n)^{25}$Mg rates, as demonstrated by Fig. 2. For these same rates, note that $^{80}$Kr is not overproduced either in some of the calculations of Goriely & Mowlavi [9] which predict high yields of the other light s-nuclides.

Note that a discrepancy, if any, between the observed Ba overabundance in the SN1987A ejecta and the model predictions could be cured in a natural way by increasing the adopted $^{22}$Ne$(\alpha,n)^{25}$Mg rate within a range compatible with the NACRE data [7].

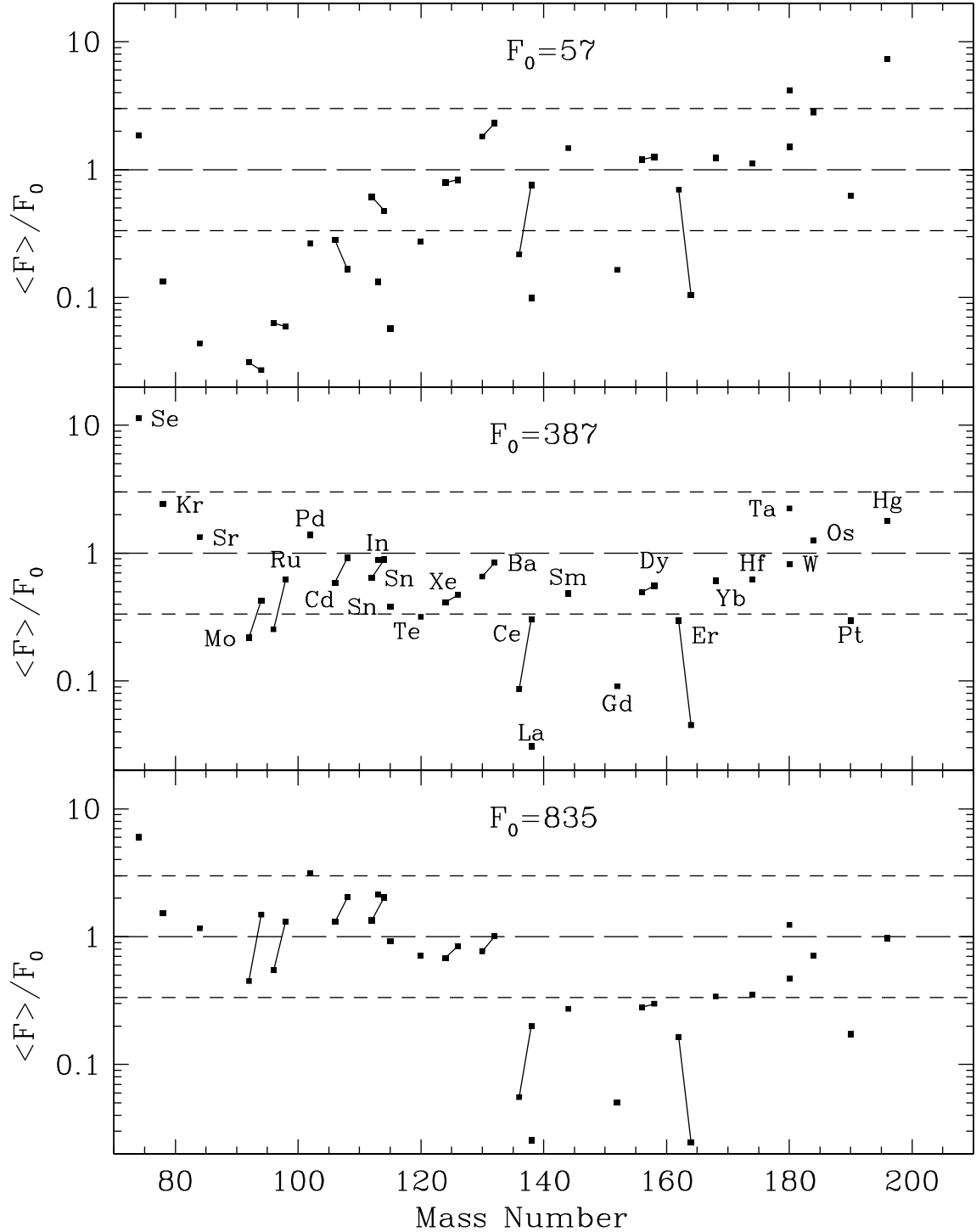

Figure 3. Values of $\langle F_i \rangle / F_0$ and of $F_0$ derived from the seed abundances calculated with the $^{22}\text{Ne}(\alpha, n)^{25}\text{Mg}$ rate R$_1$ (upper panel), R$_3$ (middle panel) and R$_5$ (lower panel).

## 4. A NON-EXOTIC SOLUTION TO THE Mo-Ru PROBLEM

The various seed abundances of Fig. 2 are used to compute the production of the p-nuclides in the PPLs of the 25 M$_\odot$ star, model 85 (as in RAHPN, for the $A \leq 40$ species, the initial abundances in the PPLs are taken from the detailed stellar models). Figure 3 shows the normalised p-nuclide overproduction factors derived from the seed abundance distributions calculated with the $^{22}$Ne$(\alpha,n)^{25}$Mg rates R$_1$, R$_3$ and R$_5$. Changes in the shape of the p-nuclide abundance distribution are clearly noticeable, at least for $A \lesssim 120$. The use of R$_1$ leads to a more or less substantial underproduction of not only $^{92}$Mo, $^{94}$Mo, $^{96}$Ru and $^{98}$Ru, but also of $^{78}$Kr and $^{84}$Sr, which was not predicted in previous calculations. This new feature directly relates to the larger abundances around $A \approx 80$ used by RAHPN (dashed curve in Fig. 2), in contrast to the much flatter seed distribution obtained with R$_1$. This Kr-Sr-Mo-Ru trough is gradually reduced, and in fact essentially disappears, for $^{22}$Ne$(\alpha,n)^{25}$Mg rates of the order or in excess of R$_3$. In these very same conditions, $\langle F_i \rangle / F_0$ for $^{113}$In and $^{115}$Sn comes much closer to unity as well. It has to be noticed that this situation does not result from a stronger production of these two nuclides by the p-process, but instead from their increased initial abundances associated with a more efficient s-process when going from R$_1$ to R$_5$. In contrast, the $\langle F_i \rangle / F_0$ pattern does not depend on the adopted $^{22}$Ne$(\alpha,n)^{25}$Mg rate for $A \gtrsim 140$. This is expected from a mere inspection of the s-nuclide seed distributions displayed in Fig. 2. In particular, $^{152}$Gd and $^{164}$Er remain underproduced. This cannot be considered as an embarassment as these two nuclides can emerge from the s-process in low- or intermediate-mass stars.

In addition, the overall efficiency of the p-nuclide production substantially increases with increasing $^{22}$Ne burning rates. More specifically, $F_0$ is multiplied by a factor of about 15 when going from R$_1$ to R$_5$. This could largely ease, and even solve, the problem of the relative underproduction of the p-nuclides with respect to oxygen identified by RAHPN. For their considered 25 M$_\odot$, model 85, star, they obtain $F_0 = 130$ and report a value of 4.4 for the ratio of the oxygen to p-process yields. This value would come close to unity for $^{22}$Ne$(\alpha,n)^{25}$Mg rates in the vicinity of R$_3$-R$_4$, as the p-nuclides would be about 3 to 6 times more produced than in RAHPN.

## 5. CONCLUSIONS

This work makes plausible that the long-standing puzzle of the underproduction with respect to solar of the p-isotopes of Mo and Ru in SNII explosions could be quite naturally solved by just assuming an increase of the $^{22}$Ne$(\alpha,n)^{25}$Mg rate over its 'nominal' value. More specifically, this could be achieved by multiplying the NACRE 'adopted' rate by factors of about 10 to 50 in the temperature range at which the s-process typically develops during core He burning in massive stars. These factors are well within the uncertainties reported by NACRE. As an important bonus, this increased rate would also largely avoid (i) the underproduction of $^{78}$Kr and of $^{84}$Sr which we predict here for the first time to be concomitant to the light Mo and Ru one, (ii) the too low production of $^{113}$In and $^{115}$Sn, and (iii) the overall underproduction of the p-nuclides with respect to oxygen noted by RAHPN. In direct relation with an increased $^{22}$Ne$(\alpha,n)^{25}$Mg rate, more s-process Ba could also be ejected by SNII events.

This array of pleasing features has of course not to be viewed as a proof of the validity of the assumption that the true $^{22}\text{Ne}(\alpha,\text{n})^{25}\text{Mg}$ rate is higher than usually thought. It may just be a hint that there might be ways around exotic solutions. This conclusion applies at least if one relies on the simplistic (and the only ones to be available for our purpose) supernova models used here and in previous p-process calculations (see RAHPN et references therein), as well as in a myriad of other explosive nucleosynthesis calculations.